\documentclass[twocolumn,english,aps,prl,notitlepage,nofootinbib,preprintnumbers,superscriptaddress]{revtex4-1}
\usepackage[T1]{fontenc}
\usepackage[latin9]{inputenc}
\usepackage{amstext}
\usepackage{esint}

\makeatletter
\usepackage{babel}
\def\Mpl{M_{\mathrm P}}

\makeatother

\usepackage{babel}
\begin{document}

\preprint{YITP-15-48, IPMU15-0081}

\title{Minimal theory of massive gravity}

\author{Antonio De Felice}

\affiliation{Yukawa Institute for Theoretical Physics, Kyoto University, 606-8502,
Kyoto, Japan}

\author{Shinji Mukohyama}

\affiliation{Yukawa Institute for Theoretical Physics, Kyoto University, 606-8502,
Kyoto, Japan}

\affiliation{Kavli Institute for the Physics and Mathematics of the Universe (WPI),
The University of Tokyo, 277-8583, Chiba, Japan}

\begin{abstract}
  We propose a new theory of massive gravity with only two propagating
  degrees of freedom. While the homogeneous and isotropic background
  cosmology and the tensor linear perturbations around it are
  described by exactly the same equations as those in the de
  Rham-Gabadadze-Tolley (dRGT) massive gravity, the scalar and vector
  gravitational degrees of freedom are absent in the new theory at the
  fully nonlinear level. Hence the new theory provides a stable
  nonlinear completion of the self-accelerating cosmological solution
  that was originally found in the dRGT theory. The cosmological
  solution in the other branch, often called the normal branch, is
  also rendered stable in the new theory and, for the first time,
  makes it possible to realize an effective equation-of-state
  parameter different from (either larger or smaller than) $-1$
  without introducing any extra degrees of freedom.
\end{abstract}
\maketitle

{\bf Introduction.---} Since the seminal work by Fierz and
Pauli~\cite{Fierz:1939ix}, especially in the recent years, much
theoretical effort in cosmology has been put in order to develop
theories of massive gravity~\cite{deRham:2010ik,deRham:2010kj}. These
theories were indeed able to introduce, at non-linear level, the
desired five modes necessary to describe a massive graviton in a
Lorentz invariant way. In other words, these theories are free from
the so called Boulware-Deser ghost~\cite{Boulware:1973my}, which had
been thought to plague any theories of massive gravity. Together with
this first success, much work came in order to see whether these same
theories could be viable. Unfortunately, these theories suffer from
instability on some key backgrounds, such as the
Friedmann-Lema\^{i}tre-Robertson-Walker (FLRW)
universe~\cite{DeFelice:2012mx}. In this regard several attempts have
been analyzed to find stable cosmological solutions in massive
gravity: 1) abandoning the homogeneity and/or the isotropy of
cosmological models; 2) changing the theory by introducing new fields
interacting with gravity in a way as to save the theory. It proved
difficult, even with these attempts, to find a theory of massive
gravity with a theoretically consistent and experimentally viable
cosmology.

In this letter, we present a new theory of massive gravity which modifies
general relativity in a minimal way. We will perform this by looking
for a theory with only two tensor modes, which are massive. This will
make FLRW backgrounds (including de Sitter) stable and viable as in
the standard cosmology. Indeed the tensor modes of the gravity sector
will be massive, whereas there are no scalar and vector propagating
modes in the gravity sector. In order to achieve this goal we will
not impose the Lorentz symmetry, so that a massive graviton does not
need to have five degrees of freedom any longer.

{\bf Precursor Theory.---} In order to define the theory we will make
use of the lapse $N$, shift $N^{i}$, and the three-dimensional
vielbein $e^{I}{}_{j}$ as basic variables. We can then introduce the
three-dimensional metric as
\begin{equation}
\gamma_{ij}\doteq\delta_{IJ}\,e^{I}{}_{i}\,e^{J}{}_{j}\,.
\end{equation}
Hereafter, $I,J\in\{1,2,3\}$ so as $i$ and $j$. Out of the variables
introduced so far, we can build a four-dimensional vielbein as 
\begin{equation}
\left\Vert e^{\mathcal{A}}{}_{\mu}\right\Vert =\left(\begin{array}{cc}
N & \vec{0}^{T}\\
e^{I}{}_{i}N^{i} & e^{I}{}_{j}
\end{array}\right),\label{eq:vielbADM}
\end{equation}
and a four-dimensional metric as 
\begin{equation}
g_{\mu\nu}\doteq\eta_{\mathcal{AB}}\,e^{\mathcal{A}}{}_{\mu}\,e^{\mathcal{B}}{}_{\nu}\,,
\end{equation}
where $\eta_{\mathcal{AB}}$ is the Minkowski metric tensor, so that
\begin{eqnarray}
g_{00} & = & -N^{2}+\gamma_{ij}N^{i}N^{j}\,,\nonumber\\
g_{0i} & = & \gamma_{ij}N^{j}=g_{i0}\,,\quad
g_{ij} = \gamma_{ij}\,,
\end{eqnarray}
corresponding to the metric tensor written in the ADM variables. We
also introduce a non-dynamical four-dimensional vielbein built out of
a non-dynamical lapse $M$, a non-dynamical shift $M^{i}$, and a
non-dynamical three-dimensional vielbein $E^{I}{}_{j}$, as follows:
\begin{equation}
\left\Vert E^{\mathcal{A}}{}_{\mu}\right\Vert \doteq\left(\begin{array}{cc}
M & \vec{0}^{T}\\
E^{I}{}_{i}M^{i} & E^{I}{}_{j}
\end{array}\right).
\end{equation}
The four-dimensional vielbein of the form (\ref{eq:vielbADM}), often
called the ADM vielbein, has $13$ independent components, as opposed
to $16$ independent components of a completely general vielbein in
four-dimensions. The missing $3$ components are the boost parameters
that would transform the vielbein of the form (\ref{eq:vielbADM}) to a
general vielbein. Therefore, by choosing the form (\ref{eq:vielbADM})
for the vielbein, we introduce a preferred frame and thus explicitly
break the local Lorentz symmetry.

We now introduce a precursor action, which will then be used as the
starting point to define the theory:
\begin{eqnarray}
S_{\mathrm{pre}} & = & \frac{\Mpl^{2}}{2}\int d^{4}x\sqrt{-g}\,\mathcal{R}[g_{\mu\nu}]\nonumber \\
 &  & +\frac{\Mpl^{2}}{2}m^{2}\int d^{4}x\,\Biggl[\frac{c_{0}}{24}\epsilon_{\mathcal{ABCD}}\epsilon^{\alpha\beta\gamma\delta}E^{\mathcal{A}}{}_{\alpha}E^{\mathcal{B}}{}_{\beta}E^{\mathcal{C}}{}_{\gamma}E^{\mathcal{D}}{}_{\delta}\nonumber \\
 &  & {}+\frac{c_{1}}{6}\epsilon_{\mathcal{ABCD}}\epsilon^{\alpha\beta\gamma\delta}E^{\mathcal{A}}{}_{\alpha}E^{\mathcal{B}}{}_{\beta}E^{\mathcal{C}}{}_{\gamma}e^{\mathcal{D}}{}_{\delta}\nonumber \\
 &  & {}+\frac{c_{2}}{4}\epsilon_{\mathcal{ABCD}}\epsilon^{\alpha\beta\gamma\delta}E^{\mathcal{A}}{}_{\alpha}E^{\mathcal{B}}{}_{\beta}e^{\mathcal{C}}{}_{\gamma}e^{\mathcal{D}}{}_{\delta}\nonumber \\
 &  & {}+\frac{c_{3}}{6}\epsilon_{\mathcal{ABCD}}\epsilon^{\alpha\beta\gamma\delta}E^{\mathcal{A}}{}_{\alpha}e^{\mathcal{B}}{}_{\beta}e^{\mathcal{C}}{}_{\gamma}e^{\mathcal{D}}{}_{\delta}\nonumber \\
 &  & {}+\frac{c_{4}}{24}\epsilon_{\mathcal{ABCD}}\epsilon^{\alpha\beta\gamma\delta}e^{\mathcal{A}}{}_{\alpha}e^{\mathcal{B}}{}_{\beta}e^{\mathcal{C}}{}_{\gamma}e^{\mathcal{D}}{}_{\delta}\Biggr]\,,
\end{eqnarray}
where $\mathcal{R}[g_{\mu\nu}]$ is the four-dimensional Ricci scalar
for the metric $g_{\mu\nu}$ and the Levi-Civita symbol is normalized
as $\epsilon_{0123}=1=-\epsilon^{0123}$. The precursor action would be
exactly the same as that for the dRGT massive gravity if
$e^{\mathcal{A}}{}_{\alpha}$ were a general, i.e.\ totally
unconstrained, vielbein in four-dimensions. At the level of the
definition of the precursor theory, the only difference from the dRGT
theory is thus that the four-dimensional vielbein is restricted to the
form (\ref{eq:vielbADM}).

Having given the action for the precursor theory, it is
straightforward to write down its Hamiltonian. The precursor
Hamiltonian turns out to be linear in $N$ and $N^i$ and independent of
their time derivatives. One can thus safely consider $N$ and $N^{i}$
as Lagrange multipliers, and the phase space to be considered here
then consists of $9\times2=18$ variables, $e^{L}{}_{k}$ and their
conjugate momenta $\Pi^{k}{}_{L}$. The coefficients of $N$ and $N^{i}$
define primary constraints, that we denote as $-\mathcal{R}_{0}$ and
$-\mathcal{R}_{i}$, respectively. The rank of the $4\times 4$ matrix
made of Poisson brackets among them is two, leading to two secondary
constraints, which we denote as $\tilde{\mathcal{C}}_{\tau}$
($\tau=1,2$). Combined with other six (three primary and three
secondary) constraints, that we name as $\mathcal{P}^{[MN]}$ and
$Y^{[MN]}$, associated with a symmetry condition on the vielbein
$e^L{}_k$, it is deduced that the physical phase space is $18-6-6=6$
dimensional and that the number of physical degrees of freedom in the
precursor theory is three at the fully nonlinear level.

{\bf Minimal Theory.---}
While the precursor theory itself is interesting, we further proceed to 
remove one more degree of freedom to define a theory with only two 
degrees of freedom, that we call the minimal theory of massive gravity. 
From now on, we will fix the units so that $\Mpl^{2}=2$. Also, we neglect 
the entirely non-dynamical part proportional to $c_{0}$. 

The minimal theory is defined in the Hamiltonian language by imposing 
four constraints, which we denote as $\mathcal{C}_0$ and $\mathcal{C}_i$
and are defined in (\ref{eqn:def-C0Cl}) below, on the precursor theory. 
Only two combinations among these four constraints are new since the 
other two independent combinations are 
$\tilde{\mathcal{C}}_{\tau}\approx 0$ ($\tau=1,2$), that already exist in the 
precursor theory. Hence the Hamiltonian of the minimal theory is
\begin{eqnarray}
H & = & \int d^{3}x[-N\mathcal{R}_{0}-N^{i}\mathcal{R}_{i}+m^{2}M\mathcal{H}_{1}\nonumber \\
 &  & +\lambda\mathcal{C}_{0}+\lambda^{i}\mathcal{C}_{i}+\alpha_{MN}\mathcal{P}^{[MN]}+\beta_{MN}Y^{[MN]}]\,,\label{eq:Hamil}
\end{eqnarray}
where $N$, $N^i$, $\lambda$, $\lambda^{i}$, $\alpha_{MN}$
(antisymmetric) and $\beta_{MN}$ (antisymmetric) are $14$ Lagrange
multipliers. This is a constrained version of the precursor
Hamiltonian, because we have added two additional constraints. As a
consequence, on the constrained surface the Hamiltonian density
reduces only to
$H\approx H_{1}\doteq\int d^{3}xm^{2}M\mathcal{H}_{1}$. Each
constraint has a specific meaning. The following terms are all derived
from the precursor theory,
\begin{eqnarray*}
\mathcal{R}_{0} & = & \mathcal{R}_{0}^{\mathrm{GR}}-m^{2}\mathcal{H}_{0}\,,\\
\mathcal{R}_{0}^{\mathrm{GR}} & = & \sqrt{\gamma}\,R[\gamma]-\frac{1}{\sqrt{\gamma}}\left(\gamma_{nl}\gamma_{mk}-\frac{1}{2}\gamma_{nm}\gamma_{kl}\right)\pi^{nm}\pi^{kl}\,,\\
\mathcal{R}_{i} & = & \mathcal{R}_{i}^{\mathrm{GR}}=2\gamma_{ik}\mathcal{D}_{j}\pi^{kj}\,,\\
\mathcal{H}_{0} & = & \sqrt{\tilde{\gamma}}(c_{1}+c_{2}\,Y{}_{I}{}^{I})+\sqrt{\gamma}(c_{3}\,X{}_{I}{}^{I}+c_{4})\,,\\
\mathcal{H}_{1} & = & \sqrt{\tilde{\gamma}}\left[c_{1}Y{}_{I}{}^{I}+\frac{c_{2}}{2}\,(Y{}_{I}{}^{I}Y{}_{J}{}^{J}-Y{}_{I}{}^{J}Y{}_{J}{}^{I})\right]+c_{3}\sqrt{\gamma}\,,
\end{eqnarray*}
and
\begin{eqnarray*}
\mathcal{P}^{[MN]} & = & e^{M}{}_{j}\,\Pi^{j}{}_{I}\delta^{IN}-e^{N}{}_{j}\,\Pi^{j}{}_{I}\,\delta^{IM}\,,\\
Y^{[MN]} & = & \delta^{ML}\,Y_{L}{}^{N}-\delta^{NL}\,Y_{L}{}^{M}\,,
\end{eqnarray*}
out of which the precursor Hamiltonian is 
$H_{\mathrm{pre}}=\int d^{3}x[-N\mathcal{R}_{0}-N^{i}\mathcal{R}_{i}+m^{2}M\mathcal{H}_{1}+\tilde{\lambda}^{\tau}\tilde{\mathcal{C}}_{\tau}+\alpha_{MN}\mathcal{P}^{[MN]}+\beta_{MN}Y^{[MN]}]$.
Here, $\tau=1,2$, $\mathcal{D}_{j}$ is the spatial covariant derivative compatible with $\gamma_{ij}$, $\sqrt{\gamma}=\sqrt{\det\gamma_{ij}}$, $\sqrt{\tilde{\gamma}}=\sqrt{\det\tilde{\gamma}_{ij}}$, $\tilde{\gamma}_{ij}=\delta_{IJ}E^{I}{}_{i}E^{J}{}_{j}$, $\pi^{jk}=\delta^{IJ}\Pi^j{}_Ie_J{}^k$, $\Pi^j{}_I$ is the canonical momentum conjugate to $e^I{}_j$, and 
\begin{equation}
Y{}_{I}{}^{J}=E{}_{I}{}^{k}e^{J}{}_{k}\,,\quad\textrm{and	}\quad X{}_{I}{}^{J}=e{}_{I}{}^{k}E^{J}{}_{k}\,,
\end{equation}
satisfying $Y{}_{I}{}^{L}X{}_{L}{}^{J}=\delta_{I}^{J}$.

Throughout the present letter, for simplicity we adopt the unitary
gauge so that $M$, $M^{i}E{}^{I}{}_{i}$ and $E{}^{I}{}_{j}$ are
only functions of the coordinates. This makes $\mathcal{H}_{0}$ and
$\mathcal{H}_{1}$ explicitly time-dependent. The remaining constraints, 
$\mathcal{C}_0$ and $\mathcal{C}_i$, are then defined as
\begin{equation}
\mathcal{C}_{0}\doteq\{\mathcal{R}_{0},H_{1}\}+\frac{\partial\mathcal{R}_{0}}{\partial t}\,,\quad\mathcal{C}_{i}\doteq\{\mathcal{R}_{i},H_{1}\}\,.
\label{eqn:def-C0Cl}
\end{equation}
The two constraints $\tilde{\mathcal{C}}_{\tau}\approx 0$ ($\tau=1,2$)
in the precursor theory can be written as linear combinations of these
four constraints. Therefore, only the remaining two combinations are
new. In other words, the minimal theory is defined by adding two
additional constraints to the precursor theory. The set of two new
constraints removes one degree of freedom from the precursor
theory. Since the precursor theory has three degrees of freedom, this
means that the minimal theory has only two degrees of freedom.

Rigorously speaking, what we have proved here is that there are enough
number of constraints, meaning the inequality,
$(\mbox{number of d.o.f.})\leq 2$, holds. One might in fact worry that
the consistency of the additional two constraints with time evolution
might lead to further secondary constraints, overconstraining the
theory. This is not the case since, as we shall see explicitly in the
next section, there are two (and only two) propagating degrees of
freedom around cosmological backgrounds, meaning another inequality,
$(\mbox{number of d.o.f.})\geq 2$, holds. By combining the two
inequalities, we thus have the equality,
$(\mbox{number of d.o.f.})=2$, holds. It is also possible to prove the
absence of further secondary constraints, and thus the presence of two
physical degrees of freedom, in a more formal way by calculating the
determinant of the $14\times 14$ matrix-operator made of Poisson
brackets among all the fourteen constraints~\cite{DeFelice:2015moy}.

Having defined the minimal theory with only two degrees of freedom by
its Hamiltonian, it is straightforward to calculate the corresponding
action via a Legendre transformation.  It should be noticed that since
in the constraints, e.g.\ $\mathcal{C}_0$, the canonical momenta are
present, on integrating out, e.g.\ the auxiliary field $\lambda$, the
resulting Lagrangian (or Hamiltonian) would acquire a structure, which
would have a kinetic structure for ${\dot\gamma}_{ij}$ essentially
different from the Einstein-Hilbert one. While this modification is
essential for the absence of helicity-0 and -1 gravitational modes,
the modification to the kinetic term for the gravitational waves is
suppressed by $m^2/\Mpl^2$ and thus is negligible. A nontrivial point
to be noticed in this regard is that the relation between the
canonical momenta $\Pi^j{}_I$ and the time derivative of the spatial
vielbein $\dot{e}^I{}_j$ is modified due to the additional terms in
the Hamiltonian~\cite{DeFelice:2015moy}. One can analyze the behavior of the
theory, e.g.\ the cosmological evolution, by using either the
Hamiltonian or equivalently the Lagrangian.

{\bf Cosmology and phenomenology.---} Let us consider a simple case to
show the behavior of the theory, namely a general homogeneous and
isotropic (FLRW) cosmological background with flat spatial metric,
driven by the graviton mass term and matter fields minimally coupled
to the metric $g_{\mu\nu}$. It is rather easy to see that on this
background, $\mathcal{C}_i\approx0$ are trivial (because of
homogeneity of the background) and $\mathcal{C}_0\approx0$ leads to
\begin{equation}
(c_{3}+2c_{2}X+c_{1}X^{2})(\dot{X}+NHX-MH)=0\,, \label{eqn:FLRW-Bianchi}
\end{equation}
where $X\doteq\tilde{a}/a$ is the ratio of the scale factors of the
three-dimensional metrics $\tilde{\gamma}_{ij}$ and $\gamma_{ij}$
respectively, and $H$ is the Hubble expansion rate (not the
Hamiltonian), i.e.\ $H=\dot{a}/(aN)$. This is exactly the same as the
well-known constraint equation obtained from the Bianchi identity in
the dRGT theory. Two branches of solutions thus exist, corresponding
to the two factors of the left hand side of
(\ref{eqn:FLRW-Bianchi}). The self-accelerating branch is defined by
those values of the constant $X$ which satisfy
\begin{equation}
X=X_{\pm}\doteq\frac{-c_{2}\pm\sqrt{c_{2}^{2}-c_{1}c_{3}}}{c_{1}}\,,
 \label{eqn:self-accelerating-branch}
\end{equation}
while the normal branch is defined by setting $\dot{X}+NHX-MH=0$. In order to determine the Lagrange multiplier $\lambda$, we demand that 
\[
\frac{d\mathcal{R}_{0}}{dt}=\{\mathcal{R}_{0},H\}+\frac{\partial\mathcal{R}_{0}}{\partial t}=\mathcal{C}_{0}+\int d^3y\lambda\{\mathcal{R}_{0},\mathcal{C}_{0}(y)\}+\dots\,,
\]
should vanish, where $\{\mathcal{R}_{0},\mathcal{C}_{0}\}\neq0$ and
the neglected part vanishes because of the symmetry of the
background. On requiring $\frac{d\mathcal{R}_{0}}{dt}\approx0$ and
imposing $\mathcal{C}_0\approx0$, we then find that $\lambda\approx0$
on the background. The remaining independent equation is, after
re-inserting units,
\begin{equation}
3\Mpl^2H^{2}=\frac{m^{2}\Mpl^2}{2}\left(c_{4}+3c_{3}X+3c_{2}X^{2}+c_{1}X^{3}\right)
 + \rho, \label{eqn:Friedmann-eq}
\end{equation}
where $\rho$ is the energy density of matter minimally coupled to the
metric $g_{\mu\nu}$. This is exactly the same as the Friedmann
equation in the dRGT theory. There is no other independent equation
for the background, essentially because both the Bianchi identity and
the constraint $\mathcal{C}_0\approx0$ lead to the same equation
(\ref{eqn:FLRW-Bianchi}). Needless to say, when we constructed the
minimal theory in the previous section, we carefully chose
$\mathcal{C}_0$ so that this is the case.

While the homogeneous and isotropic cosmological background solutions
are exactly the same as those in the dRGT theory, perturbations around
them behave completely differently. It is known that, in the standard
dRGT theory, all homogeneous and isotropic background in both branches
are plagued by ghosts, either in the helicity-$0$ or helicity-$1$
sector, and thus unstable~\cite{DeFelice:2012mx}. On the contrary, in
the new theory there is no physical degree of freedom in the
helicity-$0$ or helicity-$1$ sector (namely no scalar or vector modes,
according to the standard 1+3 decomposition of the perturbation for
the metric tensor) and thus those (would-be) ghosts are absent,
rendering the cosmological background absolutely stable. In the
minimal theory, the only existing two propagating modes reduce to the
tensor modes, whose quadratic action can be written as
\begin{equation}
S=\frac{\Mpl^{2}}{8}\sum_{\lambda={+},{\times}}\int d^{4}xNa^{3}\left[\frac{\dot{h}_{\lambda}^{2}}{N^{2}}-\frac{(\partial h_{\lambda})^{2}}{a^{2}}-\mu^{2}h_{\lambda}^{2}\right],
\end{equation}
where 
\begin{equation}
\mu^{2}\doteq\frac{1}{2}\,m^{2}\,X\,
 \left[(c_2X+c_3)+(c_1X+c_2)\frac{M}{N}\right], 
\end{equation}
gives the mass $\mu$ of the tensor modes on this background in both
branches, which is in general different from the mass parameter $m$ in
the action. This provides a proof of our previous claim that the
theory is not overconstrained. Thus, not only the background equation
of motion but also the quadratic action for tensor perturbations are
exactly the same as those in the dRGT
theory~\cite{Gumrukcuoglu:2011ew,Gumrukcuoglu:2011zh}. On the other
hand, for scalar and vector perturbations, there is no additional
modes stemming from the gravity sector. This is consistent with what
we have found in the previous section, namely the fact that the number
of physical degrees of freedom in the gravity sector is only two. (We
shall discuss each branch later in this section.) Since the only two
physical modes from the gravity sector coincide with the tensor modes,
the cosmological background is stable, provided that $\mu^{2}>0$ and
that the matter sector is stable (except for the standard Jeans
instability that drives the structure formation). In particular, the
theory automatically avoids the nonlinear ghost instability found in
\cite{DeFelice:2012mx} and the classical Higuchi
ghost~\cite{Higuchi:1986py}. This feature of the minimal theory is in
a sharp contrast to the dRGT theory.

In the minimal theory, the constraints $\mathcal{C}_0$ and
$\mathcal{C}_i$ play key roles in eliminating the unwanted
helicity-$0$ and helicity-$1$ gravitational modes. Actually, we have
chosen $\mathcal{C}_0$ and $\mathcal{C}_i$ so that they always contain
the two constraints $\tilde{\mathcal{C}}_{\tau}$ ($\tau=1,2$) in the
precursor theory as well as two additional constraints. Moreover, as
stressed just after (\ref{eqn:Friedmann-eq}), $\mathcal{C}_0\approx0$
and the Bianchi identity result in the same equation for the
homogeneous and isotropic background. These rather non-trivial
properties uniquely characterize $\mathcal{C}_0$ and $\mathcal{C}_i$.

Phenomenology in the self-accelerating branch of the minimal theory is
almost the same as the standard $\Lambda$CDM cosmology. For the
background evolution, since $X$ is set to be a constant by
(\ref{eqn:self-accelerating-branch}), the graviton mass term in
(\ref{eqn:Friedmann-eq}) behaves as an effective cosmological
constant, that can drive the acceleration of the cosmic expansion even
without the genuine cosmological constant. Scalar and vector
perturbations behave in exactly the same way as in the standard
$\Lambda$CDM cosmology. Only the tensor perturbations are modified by
the graviton mass term, as described above. Therefore, the
self-accelerating branch of the minimal theory leads to absolutely
stable and phenomenologically viable cosmology.

The normal branch of the minimal theory is also interesting. This
branch is defined by $\dot{X}+NHX-MH=0$, and thus in this branch $X$
is not a constant in general, making the contribution of the graviton
mass term to the Friedmann equation (\ref{eqn:Friedmann-eq})
dynamical. It is possible to find regimes of parameters in which
$\mu^2$ is positive and the effective equation-of-state parameter of
the graviton mass term is either larger or smaller than $-1$. This is
quite remarkable both theoretically and observationally. It is
commonly believed that the effective equation-of-state different from
$-1$ would imply the existence of extra helicity-$0$ (and possibly
helicity-$1$) degree(s) of freedom in either the dark sector or the
gravity sector. Indeed, this is one of the main motivations for
numerous dark energy surveys in the world. On the contrary, in the
minimal theory of massive gravity there is no extra degree of freedom,
and yet the cosmology with the effective equation-of-state different
from $-1$ is possible and stable.

In the absence of extra gravitational degrees of freedom, in either branch of solutions, the minimal theory is not constrained by fifth force experiments and thus the bound on the graviton mass is relatively weak. We do not even need screen mechanisms such as Vainshtein's one~\cite{Vainshtein:1972sx}. The strongest bound, 
\begin{equation}
\mu_{{\rm s}}\lesssim10^{-5}\,\text{Hz},
\end{equation}
comes from modification of the emission rate of the gravitational
waves from binary pulsars, which is of order $O(\mu_{{\rm s}}^{2}/\omega^{2})$~\cite{Finn:2001qi}.
Here, $\mu_{{\rm s}}$ is the mass of gravitational waves at the source
position and $\omega$ is the characteristic frequency of the system.

The theory opens up new possibilities for gravitational phenomenology.
One such example is possible appearance of a sharp peak in the stochastic
gravitational wave spectrum~\cite{Gumrukcuoglu:2012wt}.

{\bf Conclusion.---} We propose here a minimal theory of massive
gravity, where only two physical modes in the gravity sector
propagate, the gravitational waves, which become massive. This theory,
contrary to the dRGT theory, allows stable homogeneous and isotropic
(FLRW) cosmologies to exist. In particular, the self-accelerating
cosmological solution, that was originally found in the dRGT theory,
is now rendered stable in this theory at the fully nonlinear
level. Furthermore, we expect that its phenomenology will be closer to
General Relativity, as only the tensor modes are dynamical, as in
General Relativity. The phenomenology of gravitational waves differs
from the one in General Relativity, as they possess a non-zero mass,
giving them a different propagation dynamics (e.g.\ the propagation
speed for modes with low frequencies). The normal branch cosmological
solution is also rendered stable in the minimal theory and allows the
effective equation-of-state parameter to differ from (either larger or
smaller than) $-1$. As far as the authors know, this is the very first
example in which the effective equation-of-state parameter is made
different from $-1$ without introducing any extra degrees of
freedom. This potentially has rather strong impact on various dark
energy surveys in the world.

It is possible to extend this theory to a bigravity theory with only
$4$ physical degrees of freedom, by promoting $M$, $M^{k}E^{L}{}_{k}$,
and $E^{L}{}_{k}$ to be dynamical. Two of the four are massless
graviton degrees and the remaining two are massive graviton
degrees. It is also straightforward to generalize it to a
multi-gravity setup. We hope to extend this method to other cases for
which unwanted degrees of freedom can be consistently removed via
well-imposed constraints.

Even within the context of single graviton, there are many
possibilities for extension of the minimal theory proposed in the
present paper.  For example, one might be tempted to combine the
minimal theory with the idea of the generalized massive
gravity~\cite{deRham:2014gla}.  Namely, by introducing St\"{u}ckelberg
fields and promoting the constant coefficients $c_{0,1,2,3,4}$ to
functions of them, the dynamics of FLRW background can be altered. It
is also possible to promote $c_{0,1,2,3,4}$ to functions of other
dynamical scalar fields as in \cite{D'Amico:2012zv,DeFelice:2013tsa}.
We leave the important discussion of the nature of the Lorentz
violations, and their consequences related to the cutoff scale of the
theory as a project to be discussed elsewhere. Here, we simply state
that classically there is no Lorentz violation in the matter sector
and that Lorentz violation via loops should be suppressed by the tiny
factor $m^2/\Mpl^2$, provided that all matter fields couple minimally
to the physical metric $g_{\mu\nu}$ at the classical level.

\begin{acknowledgements}
One of the authors (SM) was supported in part by Grant-in-Aid for
Scientific Research 24540256 and the WPI Initiative, MEXT Japan. We
thank Denis Comelli and Luigi Pilo for useful discussions.
\end{acknowledgements}

\end{document}